# Model Joins: Enabling Analytics Over Joins of Absent Big Tables

Ali Mohammadi Shanghooshabad, Peter Triantafillou
University of Warwick
{ali.mohammadi-shanghooshabad,p.triantafillou}@warwick.ac.uk

**Abstract**
This work is motivated by two key facts. First, it is highly desirable to be able to learn and perform knowledge discovery and analytics (LKD) tasks without the need to access raw-data tables. This may be due to organizations finding it increasingly frustrating and costly to manage and maintain ever-growing tables, or for privacy reasons. Hence, compact models can be developed from the raw data and used instead of the tables. Second, oftentimes, LKD tasks are to be performed on a (potentially very large) table which is itself the result of joining separate (potentially very large) relational tables. But how can one do this, when the individual to-be-joined tables are absent?

Here, we pose the following fundamental questions: Q1: How can one "join models" of (absent/deleted) tables or "join models with other tables" in a way that enables LKD as if it were performed on the join of the actual raw tables? Q2: What are appropriate models to use per table, in order to capture all essential variable relationships that would help address Q1 effectively? Q3: As the model join would be an approximation of the actual data join, how can one evaluate the quality of the model join result? This work puts forth a framework, *Model Join*, addressing these challenges. The framework integrates and joins the per-table models of the absent tables and generates a uniform and independent sample that is a high-quality approximation of a uniform and independent sample of the actual raw-data join. The approximation stems from the models, but not from the *Model Join* framework. The sample obtained by the *Model Join* can be used to perform LKD downstream tasks, such as approximate query processing, classification, clustering, regression, association rule mining, visualization, and so on.

To our knowledge, this is the first work with this agenda and solutions. Detailed experiments with TPC-DS data and synthetic data showcase *Model Join*'s usefulness.

## 1 INTRODUCTION

**Problem.** The drivers for this work stem from ever-increasing tables in the era of big data (and the desire to delete them to avoid challenges of storage and maintenance costs), and/or formidable challenges imposed by privacy regulations in sharing data, and/or operating in federated/distributed environments where raw tables are not available to all, or too large to communicate between network nodes. Each of these drivers leads to operating with unavailable tables. It is thus becoming increasingly desirable to be able to replace tables with learned models, which retain essential relationships between the table's variables. We refer to such tables as "absent" to include deleted, or in general unavailable tables for any reason. However, one often needs to perform LKD tasks on the join result of tables. If tables are unavailable, how can one perform downstream LKD tasks on their join result? How can one use the per-table learned models to do so?

We wish to learn a generative model over absent tables to generate a uniform and independent sample of the join result for a model join query. A query is a model join query in which at least one of the to-be-joined tables is absent. The sample enables further 'downstream' LKD tasks, such as clustering analysis, classification, regression analysis, approximate query answering, visualization, etc). (In fact, an increasing number of LKD models in the literature are trained over such samples [8, 16, 17, 30, 31]).

Consider models $\mathcal{M}_0$, $\mathcal{M}_1$ and $\mathcal{M}_2$ over absent tables $\{\mathcal{D}_i\}_0^2$ which can answer probabilistic queries like $P(A)$ and $P(B|A)$, where A and B are variables (attributes) in the same table. $\{\mathcal{M}_i\}_0^2$ can be different types of models. *Model Join* uses $\{\mathcal{M}_i\}_0^2$ to generate (a high-quality approximation of) the join result that would have been obtained with joining $\{\mathcal{D}_i\}_0^2$.

The *"approximation"* (error) stems from the use of the per-table models instead of the actual tables. The *Model Join* framework itself does not introduce any extra error to the uniformity of the sample.

**Baselines.** Note that one cannot simply join the underlying tables as the assumption is that tables are absent. Also one cannot simply 'chain' models in a row to generate a uniform sample of the join result. Such sample will be statistically poor. Consider the following tables and attributes: $D_0(A, B)$, $D_1(B, C)$, and $D_2(C, D)$. Suppose models $\mathcal{M}_0$, $\mathcal{M}_1$, and $\mathcal{M}_2$ and $\mathcal{M}_3$ learn $P(A)$, $P(B|A)$, $P(C|B)$, and $P(D|C)$ on the tables $\{\mathcal{D}_i\}_0^2$. If we use those models in a row, they will generate a uniform sample of $P(A) \cdot P(B|A) \cdot P(C|B) \cdot P(D|C)$, which is not equal to the desired joint distribution $P(A, B, C, D)$ (as per the chain rule). That is because $P(A), P(B|A), P(C|B)$, and $P(D|C)$ are probabilities over the single tables, but not the probabilities over their joint distribution.

If we have the (models for) conditional and un-conditional probabilities in hand, one possible solution would be to create a uniform sample of each table $S(D_i)$ using the learned per-table models and then join the resulting samples. However, $S(D_0) \bowtie S(D_1) \bowtie ... \bowtie S(D_{n-1})$ is not equal to $S(D_0 \bowtie D_1 \bowtie ... \bowtie D_n)$: It is well-known [2, 4, 25] that this would produce a sample of very poor quality (i.e., not uniform). High quality samples in this setting are defined as those observing the qualities of uniformity and independence.

Therefore, we will introduce a novel framework for learning the join result by "joining models". Our solution will facilitate downstream LKD tasks on the join result. It will be generative, able to generate a uniform and independent sample of arbitrary size of the join result. The sample will serve as a *universal solution* for all downstream LKD tasks. The proposed solution will be, by nature, approximate in the sense that the generated uniform sample will be an approximation of a uniform sample of the join result. We will prove that, if the models are exact, the generated join samples will also be exact. In other words, the framework's model joining does not add any extra error – it will use exact inference algorithms.

**Why models but not data?** There are several reasons that we need to use models instead of tables. Nowadays big data operators are faced with formidable challenges managing massive tables (e.g., historical data, IoT data, monitoring/telemetry data, etc.). Hence, a strong requirement has emerged to be able to delete data, but without losing the ability to re-obtain the key information of the deleted data[11, 21]. Learning models and deleting the data is one of the options to answer the key questions about the data without storing the data. Models like [8, 17, 32] are some successful models over relational data. In fact, the major funder of this research is a leading communications infrastructure company, inundated with ever-growing telemetry/monitoring data spread across tables reporting alarms, faults, tickets, and KPI data, such as CPU and network bandwidth utilization, faults, customer SLAs and achieved performance, etc., needing to be able to perform LKD tasks over their joins and needing to delete the massive tables.

Another reason to not make raw data/tables available is that one may not be allowed access to data tables because of privacy reasons [6]. Modern privacy regulations like Europe's GDPR and California's CCPA force data owners to employ strict privacy preserving approaches when sharing data with others. Those regulations led to emerging differential privacy [1, 6, 7].

Likewise, for cases of federated learning, where one may have access to several models (coming from several data sources), but not to the underlying localized data [13, 14].

**What kind of per-table models can be joined?** In order to generate a uniform sample of the model join result, our framework needs to have the frequencies of Join Attributes (JAs) per table and conditional frequencies of a JA given another JA if a table has two JAs. The models also should provide the conditional probabilities of non-JAs (those attributes/variables that are not JAs) conditioned on JAs in the same table. As some examples, [8, 17, 32] can provide those statistics by using only one model per table. Some of the models like [8, 9] can learn the whole distribution of a table and can answer any conditional probabilities. Note all said statistics are for single tables and we do not need to have the dependencies among several tables in the models. The framework will find inter-tables dependencies using an efficient and exact approach based on dynamic programming and Probabilistic Graphical Models. The to-be-joined models could be from different types like [8, 9, 17, 32].

**Contributions.** This work contributes:

- A new framework *Model Join*, which essentially can join the models (which have replaced tables), instead of tables without adding any extra error.
- A formulation of the problem and solution leveraging PGMs thereby creating a framework whereby per-table learned models can be 'plugged in' the PGM graph.
- A tweaked Variable Elimination Algorithm to not calculate marginals, as per usual, but instead to compute all statistics necessary to build a sample generator. The end result is a generative model, able to generate an arbitrary number of high-quality approximations of the data tuples in the join result.
- If we need to conduct a model join query involving multiple models and tables, *Model Join* learns all of the necessary models on those available tables with a new efficient and accurate method then join the models. Each per-table model entails a novel blending of embeddings, clusterings, and feed forward Neural Networks (NNs).
- A detailed experimental evaluation, analyzing this new problem and quantifying the quality of the produced sample, the efficiency of *Model Join*, the dependencies on the per-table model errors, and its appropriateness for downstream analytics tasks.

The code and documentation is available at: https://github.com/shanghoosh1/ModelJoin

Section 2 provides the key background on graphical models. Section 3 presents the *Model Join* framework. Section 4 shows how the *Model Join* framework builds its own models if the data exists. Section 5 contains the experimental evaluation of the framework and its components. Section 6 discusses related work. Section 7 concludes the work.

## 2 BACKGROUND: PROBABILISTIC GRAPHICAL MODELS (PGMs)

A PGM represents a distribution in a factorized way with a graph $G(V, E)$ and some rules $\mathcal{M}$, where $V$ is a set of vertices (a.k.a. variables or nodes) and $E$ is a set of edges among the nodes. For example, a Markov Random Field (MRF) is a type of PGM having an undirected graph $G$ and three (Pairwise, Local and Global) Markov properties as $\mathcal{M}$ which are defined by the concept of *conditionally independence*. Two variables $A$ and $B$ are conditionally independent given $C$ if $P(A, B|C) = P(A|C) \times P(B|C)$.

Given $G(V, E)$, the Pairwise Markov property says that two non-adjacent variables are conditionally independent if all other variables are observed. The Local Markov property expresses if the neighboring variables of a variable are fully observed that variable is conditionally independent from all other variables. Finally, the Global Markov property states that two sets of variables are mutually independent if a separating set of variables are observed.

With PGMs, people usually use Variable Elimination Algorithm (VEA) which is an exact inference algorithm based on dynamic programming, and it is employed to calculate the marginals over the factorized distribution. For example, to calculate the marginal of $X$ in the factorized distribution $\phi(X, Y) \times \phi(Y, Z)$ with an elimination order $\sigma = \{Z, Y\}$, first, VEA sums out $Z$ from $\phi(Y, Z)$ and calculates a factor for $Y$ as $\phi(Y)$ then it sums out $Y$ from $\phi(X, Y) \times \phi(Y)$. Obviously, summing out the variables one by one from a factorized distribution (like $\phi(X, Y) \times \phi(Y, Z)$) is faster than summing out the variables ($Y$ and $Z$) from the larger un-factorized distribution $p(X, Y, Z)$. Note, $\phi(X, Y) \times \phi(Y, Z) = p(X, Y, Z)$.

## 3 THE MODEL JOIN FRAMEWORK

Figure 1 depicts an overview of *Model Join*. The input for the *Model Join* is an acyclic model join query with the available models and tables plus some meta data showing which variables are in which tables, and which models are for which tables. Recall that a query is a model join query in which at least one of the to-be-joined tables is absent. As mentioned, currently there is no solution available in the literature to perform this model join. *Model Join* creates a PGM graph based on the query and metadata before running a model join query. In general, there are two components to creating PGMs: a qualitative component that includes a graph with nodes and edges, and a quantitative component that contains the real dependencies between the nodes. In our case, once the PGM graph is obtained, the dependencies in its edges (quantitative component) come from the models. If we do not have the model for an edge, we build the model on its corresponding available table (as will be explained in Section 4). Once the PGM for the given query is ready, an inference algorithm (Variable Elimination Algorithm (VEA)) is applied to the PGM starting from the leaves to the root so that it provides all needed statistics to generate the uniform and independent sample of the join result (will be explained in Section 3.1). Then ancestral sampling is employed to generate the samples as the universal solution for downstream tasks by using those statistics derived from the inference (will be explained in Section 3.2).

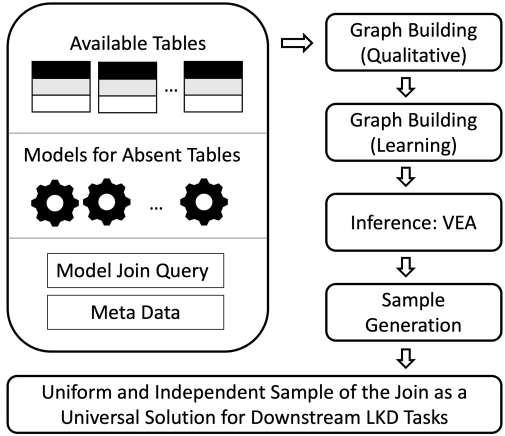

Figure 1: Overview of *Model Join*

| D1 | | D2 | | | D3 | | D4 | |
|---|---|---|---|---|---|---|---|---|
| **A** | **B** | **B** | **C** | **D** | **D** | **E** | **E** | **F** |
| a1 | b1 | b1 | c1 | d1 | d2 | e1 | e1 | f1 |
| a1 | b2 | b1 | c2 | d1 | d2 | e2 | e2 | f1 |
| a2 | b3 | b3 | c2 | d1 | d2 | e3 | e2 | f1 |
| a3 | b3 | b3 | c2 | d2 | d1 | e3 | e2 | f2 |
| … | … | … | … | … | … | … | … | … |

Figure 2: Example relational tables

Consider the need to enable LKD tasks over the join of tables $\{D_i\}_1^4$, i.e., on the result of the SQL query

**SELECT** [ d e s i r e d a t t r i b u t e s ] **FROM** D1, D2, D3, D4
**WHERE** D1.B = D2.B **and** D2.D=D3.D **and** D3.E = D4.E

Suppose further that tables $\{D_i\}_1^4$ (e.g., in Figure 2) have been deleted (or are otherwise unavailable) and only the models $\{\mathcal{M}_i\}_1^4$ over each table are available. The goal is to join the models and generate a (uniform and independent) sample of the join result, "similar" to a same size uniform sample of the join of the actual tables. The similarity between (a sample of) the actual join result and the uniform sample (of the same size) generated by *Model Join* strongly depends on the quality of the per-table models which replace the actual tables. (*Model Join* does not add any error).

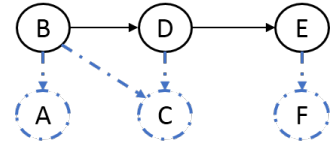

Figure 3: Join query graph on tables in Figure 2

The framework conceptually constructs a small acyclic graph employing all models. This graph has few nodes for JAs, non-JAs and edges among them if they are dependent. Specifically, if two nodes $A$, $B$ referring to table attributes, are in the same table, say $D_1$, then $A$ and $B$ are dependent and a directed edge exists between them. This edge can reflect any kind of function giving the $P(B|A)$, i.e., what we call "models". The join order specifies on which variable the condition should be defined (i.e., whether the edge $A \rightarrow B$ or $B \rightarrow A$ is drawn).

A key insight is that gathering all the per-table models in a graph as factors can provide one with the full factorized joint distribution across all models. Therefore, the graph can be thought of as a Probabilistic Graphical Model (PGM). Hence, we formulate our solution borrowing principles of PGMs (e.g., Markov properties). We develop an adapted inference algorithm over the PGM graph, which rests on incorporating and integrating the per-table models into its inference process. Please note that the models obviously will provide (only) an approximation of the relationships (e.g., conditional probabilities) among a table's attributes. However, the PGM inference itself will be exact. Therefore, as it can be proved that finding the joint distribution is exact with no additional error, if the input models are 100% accurate, the output joint distribution will in that case also be exact, error-free. With this formulation, the key issues are how to incorporate the models in PGM inference and how to derive as accurate as possible (and efficient) models for existing tables.

The JAs and the edges among them in the graph form what we refer to as the "skeleton"; all non-JAs form the non-skeleton component. The idea is that JAs and the correlations between them affect the distribution of the join result, and non-JAs do not affect the uniformity of our samples. Non-JAs are conditionally independent of the JAs of other tables, when the JAs of the same table are observed based on the Global Markov property. So, the framework focuses on the skeleton to find the statistics for the joint distribution and at the end the non-JAs of tables $D_i$ can be reached in the graph by the JAs of the same table $D_i$. Figure 3 is an example graph for the above join query on the tables from Figure 2. The black nodes and edges are the skeleton and the blue ones are the non-skeleton components.

Any node in the skeleton can be considered as the root if we change the join order (B is the root in Figure 3). The framework uses a dynamic programming algorithm to do inference on the skeleton in order to find the marginal of the root over the joint distribution of the all nodes. The inference algorithm is essentially a Variable Elimination Algorithm (VEA) [33] in PGMs. The difference is that in PGMs, people learn the factors per edges, but here the factors are already learned as per-table models, thus we use the to-be-joined models as our factors (edges in the PGM graph) when running our VEA. Another key distinction is that in PGMs, a VEA is used to calculate single marginals across a factorised distribution, whereas in our scenario, the VEA provides all the statistics required to construct the samples (like a sample generator).

At a high level, *Model Join* works as follows:

- During qualitative learning, for each attribute involved in the query, *Model Join* adds a node in the graph. If two attributes are in the same table, an edge is added between them. All attributes and egdes related to the JAs comprise the skeleton.
- During quantitative learning, *Model Join* places the available models in the corresponding edges. If the model is not available, but the table is, a model is learned on the table.
- Inference is done just once, from the leaves of the skeleton to the root, to find the marginal of the root. During inference, the framework calculates intermediate statistics which help to generate a sample of the joint distribution without running any other inference (marginalization by VEA) to achieve the full factorized joint distribution.
- Once the inference from the leaves to the root finishes, the output is a tree (including the models and the intermediate statistics) representing the factorized joint distribution. The tree can be considered as our sample generator model.
- Please note: in order to compute the full joint distribution, PGMs typically rely on a 2-phase Message Passing Protocol, where each phase is a VEA. In *Model Join* the VEA is run only once.
- A sampling algorithm (akin to Ancestral Sampling [3]) then generates a uniform sample of the joint distribution of the skeleton.
- Once the sample for the skeleton is generated, the non-skeleton variables can be added to the samples by using the models in the edges between the non-JAs and the JAs of the same table.
- The above concern acyclic joins. To deal with cyclic queries, one simply breaks cycles by eliminating edges. Then the sample for the remaining graph is generated and is rejected based on the models in the eliminated edges. Rejection sampling has already been used in [23, 34] which proved that the output from rejection sampling is a uniform sample.

## 3.1 Inference: Stats Computation with VEA

To do inference in a join graph like that of Figure 3, the framework needs the unconditional frequencies per distinct value in all JAs, and the conditional probabilities $P(X|Y)$ per $Y \rightarrow X$ edge. Note, all the probabilities can be converted to frequencies and vise versa if we have the tables' sizes, thus we need the table sizes in the metadata. In Figure 3, B, D and E and the edges among them form the skeleton of the graph. So, we need some models that provide the frequencies per B, D, and E value in the (absent) tables, and and $P(E|D)$ and $P(D|B)$ on the (absent) tables $D_3$ and $D_2$, respectively if we want to generate the variables in order of B, D and then E. The inference is carried out in reverse order E, D and then B.

Since the skeleton variables may be present in several tables, they have several frequency values, so here, lets denote the frequencies based on the table number and the position of the JAs. For example, there are two JAs in $D_3$: The left (or first) D and the right (or second) E. We denote them as $f_3^0(dv)$ and $f_3^1(dv)$ (the superscript 0 (1) denotes the first (second) JA) instead of using the node names D and E. $dv$ is for any distinct value in the JAs. Note, all the JAs are categorical. The algorithm starts from the leaves to the root and calculates the intermediate frequencies $\mathcal{F}_i$ in a deterministic and dynamic programming manner because it does not need to go through all the paths several times that it has seen already. Any $\mathcal{F}_i$ provides the frequencies of distinct values in the joint distribution of the sub-graph from node $i$ towards the leaves of the graph. In our example E is the leaf and B is the root. The inference starts from E to B and provides the frequencies of distinct values in E over table D3 ⋈ D4, the frequencies of distinct values in D over table D2 ⋈ D3 ⋈ D4 and the frequencies of distinct values in B over table D1 ⋈ D2 ⋈ D3 ⋈ D4 denoted as $\mathcal{F}_3$, $\mathcal{F}_2$ and $\mathcal{F}_1$ respectively. These frequency calculations are done in a dynamic programming way and the algorithm uses $\mathcal{F}_3$ to find $\mathcal{F}_2$, and uses $\mathcal{F}_3$ and $\mathcal{F}_2$ to find $\mathcal{F}_1$. The sum of all frequencies in $\mathcal{F}_1$ is the join size of the query which is denoted as $\mathcal{F}_0$.

In a chain join (like in our example) there is a leaf without any edge (model). So, its $\mathcal{F}_{n-1}$ can be computed by multiplying the frequencies in the underlying tables using Equation 1. For example, to find an $\mathcal{F}$ for node E (leaf), we need to multiply the frequencies in the tables D3 and D4. Recall, the tables are not available and the frequencies are provided by the per-table models.

$$\mathcal{F}_{n-1} = f_{n-1}^1(dv).f_n^0(dv), \text{for any dv } \in f_{n-1}^1 \quad (1)$$

$\mathcal{F}_{n-1}$ for our example will contain two entries (e1,1) and (e2,3) which shows the frequency of each distinct value in E over the join of D3 and D4. Note, entries with zero frequencies are ignored. To find all $\mathcal{F}_i$ in different steps of dynamic programming, Equation 2 is used. This operation continues until it reaches $\mathcal{F}_0$.

$$\mathcal{F}_i(dv) = f_i^1(dv) \times f_{i+1}^0(dv) \times \Lambda_{i+1}(dv), \text{for any dv} \in f_i^1 \quad (2)$$

where

$$\Lambda_{i+1}(dv) = \sum_{DV \in f_{i+1}^1} P_{i+1}(DV|dv) \times \frac{\mathcal{F}_{i+1}(DV)}{f_{i+1}^1(DV)} \quad (3)$$

$P_i(DV|dv)$ is the conditional probability of each distinct value $DV$ in the second JA given a specific $dv$ in the first JA, of the same table $D_i$. This conditional probability is provided by the model. Equation 2 can be considered as a step in the variable elimination process by using a sum-product operation.

In our example, once $\mathcal{F}$ for E is calculated, the $\mathcal{F}$ for D is calculated by Equation 2 which contains only one entry (d2,4). $\mathcal{F}_0$ is the join size and is calculated by summing all frequencies in $\mathcal{F}_1$.

All equations produce accurate results when accurate conditional probabilities (from the models) are available.

Once we computed all $\mathcal{F}$s, the sample generator model is ready and we have provided everything the ancestral sampling algorithm needs to generate the samples from the joint distribution.

## 3.2 Sample Generation

The join frequency tables $\mathcal{F}_i$ and the to-be-joined models together can provide the conditional probabilities using the Chow–Liu tree [5] algorithm, which we consider as our generative model. Sampling over the tree proceeds from the root to the leaves in reverse order of the variable elimination used in VEA. In our example, we can obtain $P(B)$, $P(D|B)$ and $P(E|D)$ over the joint distribution $P(B, D, E)$ and generate the uniform samples with our Algorithm 1.

The algorithm generates samples of size $n$ for JAs, one by one. In our example, it generates $n$ values for B, next for D, and then for E. For the first column of the skeleton sample $S_1$, we can simply generate samples from the frequencies in $\mathcal{F}_1$. Recall, the sum of frequencies in $\mathcal{F}_1$ is equal to the join size $\mathcal{F}_0$, so the probability of each distinct value over the joint distribution is $\frac{\mathcal{F}_1(dv)}{\sum \mathcal{F}_1}$ which can give us a uniform sample. For the rest of the skeleton columns in the sample, Equation (4) is used. For each value in $S_1$ of the skeleton, Equation (4) gives us the frequencies of all distinct values in $S_2$ over the full join result, so we can sample one value among them, and continue this process until $S_m$, where m is the number of JAs. The probabilities in $S_i$ are dependent only on the values from $S_{i-1}$. In our example, the sample for B is derived from $\frac{\mathcal{F}_1(dv)}{\sum \mathcal{F}_1}$; D values are dependent only on B values, and E values are dependent only on D values. This is because we find the $P(B)$, $P(D|B)$ and $P(E|D)$ over the joint distribution $P(B, D, E)$ (not over the single tables) by using $\mathcal{F}$s and the involved models.

Suppose $(dv_1, dv_2, ..dv_{i-1})$ have been sampled (already observed) for a single sample point. The probability of $dv_i$ over the join result is calculated by the following formula.

$$p(dv_i|dv_1, dv_2, .., dv_{i-1}) = \frac{P_i(dv_i|dv_{i-1}) \times \frac{\mathcal{F}_i(dv_i)}{f_i^1(dv_i)}}{\sum_{dv \in \mathcal{F}_i} P_i(dv|dv_{i-1}) \times \frac{\mathcal{F}_i(dv)}{f_i^1(dv)}} \quad (4)$$

The algorithm for generating samples is shown in Algorithm 1. $m$ is equal to the number the JAs, and $n$ is the sample size. $S$ is the final sample of the skeleton. Once the skeleton is generated,

---

**Algorithm 1** Generate a random sample of the skeleton

**procedure** SKELETON GENERATION($\mathcal{F}, f, m, n$)
  $S_1 \leftarrow$ randomly sample $n$ values from $\mathcal{F}_1$
  **for** i=2,3, ..., m **do**
    $S_i \leftarrow$ sample using Equation (4) for any dv $\in S_{i-1}$
    $S \leftarrow S_i$                 ▷ Append $S_i$ in $S$ column-wise
  **end for**
  **return** $S$
**end procedure**

---

then non-JAs are generated. Non-JAs are conditionally independent from the JAs in other tables, and once the JAs in the same table are given, the non-JAs become independent from other JAs. So, the models should be able to provide the probabilities for non-JAs in a table conditioned on the JAs of the same table. We generate only the non-JAs that are involved in the query (e.g. are present in the Select clause of the SQL query).

Note that the algorithm is deterministic and it should be clear that if the individual models are 100% accurate, the formulas calculate the frequencies correctly, error-free.

### 3.3 Uniformity of Generated Sample

Resting on PGM principles (Markov Properties) we prove that if the models are accurate, the samples generated by Algorithm 1 is a uniform sample of the joint distribution of all nodes.

**Theorem 1.** *If the models are accurate, the sample of the skeleton generated from Algorithm 1 is uniform.*

PROOF. The algorithm generates samples for the root by using $\frac{\mathcal{F}_1(dv)}{\sum \mathcal{F}_1}$ which clearly produces a uniform sample. This is because $\mathcal{F}_1$ is the true marginal frequencies of the root (B in our example) over the joint distribution ($P(B, D, E)$ in our example), and $\sum \mathcal{F}_1$ is equal to $\mathcal{F}_0$ which is the join size. So, the division gives us the probabilities to derive a uniform sample for the root.

To generate values for the next variable (say E in our example) after the last generated variable (D in the example), we only need the frequencies of E over the sub-graph from E to the leaves (which is provided by $\mathcal{F}$). E becomes independent from the parent of its parent (B) if its parent (D) is observed (based on the Global Markov property), so there is no need to have the frequencies of the E over the full joint distribution.

Recall that the Global Markov property states that two sets of variables are mutually independent if a separating set of variables are observed. $X_A \perp\!\!\!\perp X_B \mid X_S$ where A and B are separated by $X_S$, in other words all the paths from A to B go through S. This means that probabilities for any intermediate variable are only dependent on its offsprings to the leaves if the parent of that intermediate variable is observed. The observed parent here has the separator role in the Global Markov property. Since the join frequency table $\mathcal{F}$ for a variable X provides the frequencies of distinct values of X over the join of all the offsprings of X to the leaves, it follows that equation 4 uses the correct statistics to derive a uniform sample.

Non-JAs do not affect the uniformity of the samples at all. Adding the non-JAs also observes uniformity because non-JAs become independent from JAs in other tables if the JAs from the same table are observed (again, based on the Global Markov property). □

Nonetheless, in practice models will not be 100% accurate. What happens then? The frequencies would be calculated with some error. Since the models are inherently approximate, the output sample is a uniform-independent sample with replacement from the approximate joint distribution. The approximation is unavoidable. The framework's model joining does not add any extra error because it uses an exact inference algorithm.

## 4 LEARNING PER-TABLE MODELS

The *Model Join* framework can make its own models based on the input raw data tables. One could use a variety of learned models over relational data – for example, [8, 16, 17]. However all such models may suffer from poor accuracy (in estimating conditional probabilities) when dealing with attributes having a high number of distinct values (NDVs), as shown in [16]. On the other hand, building accurate models is very important in the model join problem because the error propagates when we are to join several models. Thus, we need models that can guarantee high accuracy. Here we present our proposal that addresses these problems. Nonetheless, the contributed framework is not dependent on any particular model and any new/future contribution in this area can be easily incorporated by *Model Join*.

For any (available) middle table in a join query, we need a model which provides the probabilities for the second JA conditioned on the first JA in that table. At this point, it is worth pointing two fundamental challenges when learning from tabular data. Namely, handling (i) non-ordinal categorical attributes and (ii) the larger Number of Distinct Values (NVDs) in attributes. The key problem with categorical attributes is that even if we use one-hot or binary encoding, we cannot learn a high-quality model over categorical attributes. This is due to the fact that the different values for categorical attributes have no meaningful relationship. The challenge with greater NDVs is that the model may have to predict a huge number of probabilities per distinct value, and learning then becomes impractical when there are large number of distinct values.

To address the first challenge, we learn embeddings per distinct value of a JA. In this way, distinct values of an attribute acquire similar embeddings, (positions in the embedded space) if they are related to similar distinct values for the other attributes. This tends to simplify the learning task and improve accuracy as relations between distinct JA values can now be detected based on how close they are in the embedded space. To address the challenge of great NDVs, after finding embeddings for distinct JA values, clustering is used on their embeddings. After clustering the second JA's values (based on their embeddings), sub-models are learned per cluster, then all sub-models are used as a single model (as will be explained shortly). Clustering makes the model more accurate because it deals with smaller learning spaces. Furthermore, it helps to achieve higher efficiency in predicting the conditional probabilities because for a given distinct value of a first JA, only a small number of the sub-models are used to predict the probabilities of the distinct values in the second JA. Moreover, this clustering can be considered as a mechanism to guarantee higher accuracy. As will show in the experiments, if the accuracy of the model is low then we can increase it by employing more clusters. To have a model with 100% accuracy, we can use $d$ number of clusters where $d$ is equal to the number of distinct pairs values in the first and second JAs. This could be materialized with a nested hash table which for a given distinct $v$ value in the first JA, it returns the probabilities per distinct value in the second JA which are related to $v$. For unconditional frequencies over JAs, we simply keep a dictionary of frequencies.

### 4.1 Embeddings

Our proposed models are conditional and also discrete in nature (as JAs are categorical). This calls to mind models used in NLP tasks, operating on individual words (which are discrete values). So, the embeddings we seek will be adapted from NLP models. Embeddings in NLP tasks yield a learned representation of words, so that the words with the same meaning have the same real-valued vector representation. These vector representations are critical in deep learning because if the input of a Neural Network (NN) model does not have meaningful distances between values, it cannot work well. Since finding good models on JAs is not simple, specially, when the JAs are keys (i.e., high NDVs), the embedding approach helps us to find more accurate models. There are many embedding approaches; embeddings like Continuous Bag-of-Words (CBOW) and Skip_Gram [20] have been highly successful for NLP tasks because of the deep linguistic theory behind them (coined distributional hypothesis). In this paper we use Skip_Gram with negative sampling [20] which is faster than the naive Skip_Gram. Notably, the way we adapt these embeddings for our task is based on the values contained in all attributes of interest. For example, in table D1 of Figure 2, distinct values $a2$ and $a3$ of attribute $A$ have some common values in $B$. Conversely, distinct values $a2$ and $a1$ do not. Therefore, the embedding vectors for $a2$ and $a3$ will be similar, whereas for $a1$ and $a2$ the embedding vectors will be different. These vector representations are used instead of the distinct values and help the final model to distinguish easily among different conditions in the input of the models. This technique significantly increases overall accuracy.

Skip_Gram is a simple NN with a single hidden layer with $N$ dimensions as embedding vectors. The goal is to learn embedding vectors. Formally, given a sequence of training words $w_1, w_2, w_3, ..., w_M$, the objective of the Skip-gram model is to maximize the average log probability

$$\frac{1}{M} \sum_{m=1}^{M} \sum_{-c \leq j \leq c, j \neq 0} \log p(w_{m+j} \mid w_m) \quad (5)$$

where $c$ is the size of the context in each word, and $M$ is the number of words in the vocabulary.

In the last layer of the Skip-Gram NN, a Softmax function is used to turn the logits to probabilities. To find $p(w_O|w_I)$ the following formula is used:

$$p(w_O|w_I) = \frac{exp(v'_{w_O}{}^\top v_{w_I})}{\sum_{w=1}^{M} exp(v'_w{}^\top v_{w_I})} \quad (6)$$

where $v_w$ and $v'_w$ are the input and output vector representations of the word $w$.

*4.1.1 Negative sampling* The Skip-gram has a large number of weights and all of them should be updated according to (maybe billions of) training data instances. Instead of updating all weights, negative sampling [20] helps to change the problem to a binary classification problem and randomly select just a small number of *negative* words and then try to distinguish between randomly chosen negative words and the current word so that there is no need to compute the similarity of one word with all other words in the corpus. The following equation is replaced with every $\log p(w_O|w_I)$ in the Skip-gram loss function Equation 5.

$$log p(v'_{w_O}{}^\top v_{w_I}) + \sum_{i=1}^{k} \mathbb{E}_{w_i \sim P_n(w)}[\sigma(-v'_{w_i}{}^\top v_{w_I})] \quad (7)$$

where there are $k$ negative samples, and where $\sigma(x) = 1/(1 + exp(x))$ and $P_n(w)$ is the noise distribution. (For more information please refer to [20].)

Mapping this to our problem, each tuple is viewed as a word sentence and the distinct values for each join attribute as the individual words. The difference between this Skip-Gram and the vanilla Skip-gram NN is that here the place of words is important (much like in [28] and [15]). For example, the distinct values in the first JA are more likely different than the distinct values in the second JA.

### 4.2 Clustering

*Model Join* incorporates a clustering approach for embeddings vectors. However, the proposed clustering here employs a twist: clustering is *based on dissimilarity*. The method is as follows: We cluster (embeddings of) distinct values in the $(i+1)_{st}$ JA based on their embedding vectors, and build different sub-models per cluster, so that each sub-model predicts a subset of distinct values in the $(i+1)_{st}$ JA, given any distinct value from the $i_{th}$ JA. Specifically: First, the distinct values in the $(i+1)_{st}$ JA are clustered to (temporary) clusters using the K-Means algorithm (actually any clustering algorithm could be used) according to the similarities of embedding vectors. Subsequently, the final clusters are created, with each one containing a (uniformly) sampled fraction of each temporary cluster (please see Figure 4). This ensures a dissimilarity among a cluster's values.

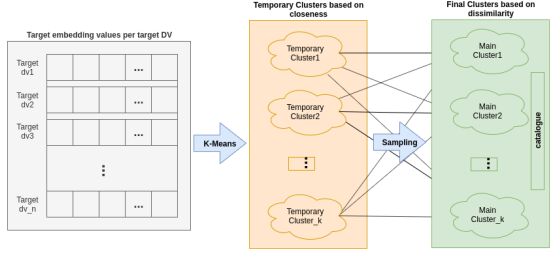

Figure 4: Clustering based on dis-similarity

This dissimilarity of values between clusters is crucial: It helps to build more accurate models per cluster. That is because when clustered data items are far from each other, modelling algorithms can distinguish among them more easily. We also store what percentage of each distinct value in the $i_th$ JA is related to the distinct values in the $(i+1)_st$ JA per cluster. This enables treating a set of sub-models (per cluster) as a single model (among all data in the clusters). For a given distinct value in the $i_th$ JA, we only call the sub-models related to that distinct value and after the prediction, we calculate the true probabilities per distinct value of the $(i+1)_st$ JA by taking the average of the probabilities.

This clustering can improve accuracy. If the accuracy of the sub-models is lacking, the number of the clusters can be increased, which will increase accuracy - as our experiments show. For a 100% accuracy, instead of models, nested hash tables can be used.

### 4.3 Conditional-Discrete Generative NNs

The final component for each per-table model are the NN models. Any "middle" table $D_i$ in the Join Query (JQ) ($D_1 \ldots D_{i-1} \bowtie_X D_i \bowtie_Y D_{i+1}, \ldots D_n$) will have two JAs (i.e., X and Y). For every single pair of these (X,Y) JAs of $D_i$, a model is sought as a *conditional generative model*. That is, to find the probabilities for the distinct values of the second JA (i.e., Y), given the value of the first JA (i.e., X). In the recent literature, one can find several such models, ranging from conditional generative adversarial models (GANs) or auto-encoders (AEs) such as [18, 19]. Although meant for continuous variables, these can be re-parameterized with Gumbel-Softmax[10] to be discrete as well. However, as the NDVs in the first JA increases, these models underperform. Furthermore, the tuning and training tasks of GANs and AEs are time- and resource-consuming. For these reasons, we turned our attention to simpler (to train, tune, and evaluate) models like probabilistic classifiers, which we adapted to use as conditional-discrete generative models. Here, we re-purpose simple feed-forward NNs, which have been used for classification with high success. Figure 5 shows the architecture of our feed-forward Conditional-Discrete Generative NN (CDG-NN) models. Since JAs are categorical, the output of the models are also categorical. Then the problem resembles multi-class classification. The input layer of the model are the embeddings of the categorical values in the first JA. After hyper-parameter tuning, we chose 5 hidden layers with the *tanh* activation function and a softmax at the end. This softmax layer provides the probabilities per distinct value in the second JA. The loss function is Negative Log Likelihood (NLL). NLL estimates the dissimilarity between the empirical distribution defined by training data and the model distribution. NLL is defined as follows:

$$\mathcal{L} = \mathbb{E}_{x \sim \hat{p}_{data}}[log\ p_{model}(x)] \qquad (8)$$

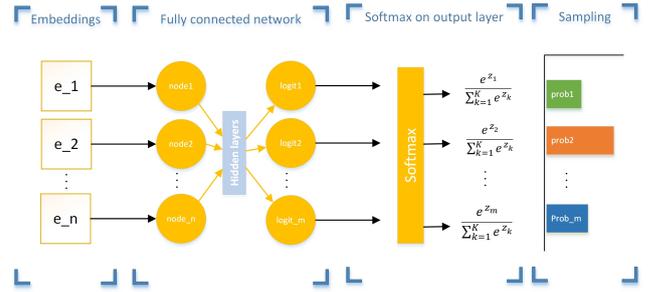

Figure 5: The CDG NN Architecture

When training, the task is like classification, but the usage of the models is slightly different. In a classification task, given an input, the softmax at the last layer yields the probabilities per output class. Then the argmax function is applied on probabilities to find the winning class. Here, instead of using argmax, *we sample using those probabilities per class*. Thus, for each given distinct value in the first JA, this finds the probability for each distinct value in the second JA. Then one can generate (as many as needed) distinct values in the second JA.

### 4.4 Evaluating a CDG NN

Consider the four main elements of the confusion matrix: True Positive (TP), True Negative (TN), False Positive (FP), and False Negative (FN). For generative models, like in classification tasks, we can also calculate all needed measures for CDG-NN models, such as recall, precision and F-score. Hence, we can suppose that any generation using a CDG-NN is a binomial task that can assume correct and wrong states. We define the confusion matrix for CDG-NN models as follows. We probabilistically sample distinct values in the output layer. Given a distinct value from the first JA: TP is the probability of correctly generating a distinct value of the second JA. FN is the probability that a distinct value of the second JA should have been generated, but has not been. FP is the probability that a distinct value of the second JA should have not been generated, but has been. Finally, TN is the probability of correctly not generating a distinct value from the second JA. For example, suppose given a *dv* from the first JA, we generate probabilities for 5 distinct values in the second JA. Said probabilities sum up to one. We also have the real probabilities from the actual table. The sum of these is also equal one. Once the F-score is calculated, if we multiply it with the number of samples which should be generated, we will get the number of correct and wrong generations. Since all TP, TN, FP and FN sum to one, precision, recall and F-score are equal. We start generating a random sample of a given size, say 10k, using 10k distinct values from the first JA, according to their frequencies. Then, one can find the actual probabilities for all distinct values of the second JA, conditioned on the distinct values from the first JA. Then, the same 10k distinct values from the first JA are input to the CDG NN model. The model generates approximate probabilities of the distinct values in the second JA. We now have both real and approximated probabilities. Per sampled distinct value from the first JA, we calculate the F-score. Since we can regard the process as a success–failure experiment, we can use the Binomial Proportion Confidence Interval (BPCI) to find error intervals. We use the Wilson score interval which is reliable even if the sample size is small or the success probability is close to 0 or 1. So, for any CDG-NN model, an average error and confidence interval can be provided.

## 5 EXPERIMENTAL EVALUATION

Our experimentation addresses five key issues: (i) quantification of key overheads; (ii) accuracy of the proposed per-table models; (iii) accuracy of the overall framework; (iv) showing that the framework's output sample is indeed a uniform sample of the *true join result*; and (v) exemplify high-quality downstream LKD tasks being performed on the framework's output.

Two sets of experiments are presented. One with very large tables to show that table sizes do not affect the framework's efficiency and only NDVs do. However, when tables are too large (expensive to join), the ground truth to measure actual errors cannot be calculated. So, in the 2nd set of experiments, smaller tables are used so the join can be computed and errors be measured.

**Metrics**. All reported times (space) are in seconds (MBs). For the quality of generated samples we use (i) the KS-test for sample uniformity and (ii) the F-score for quantifying distances from uniform-random samples of the exact join.

**Competing Aproaches**. To the best of our knowledge, this is the first solution to the stated problem. Other recent efforts like [23] and [34], can accurately and efficiently generate uniform samples over joins where the tables are not absent. However, these methods cannot handle the case where we deal with a model join query involving at least one model. So, our evaluation is conducted to analyze *Model Join*'s performance only.

### 5.1 Experimental Setup

**System configuration**. Model training uses a GPU GeForce RTX 2080 Ti, with 11 GB memory. Model joining uses a system with 64GB main memory and the E5-2660 CPU with 20, 2.60GHz cores.

**Data and queries**. In Table 1 we show statistics for the synthetic and benchmark data. The first synthetic data $SynthDB1$ has a fixed NDVs, but variant NDVs pairs (first JA, second JA). The second synthetic data $SynthDB2$ has variant NDVs in the JAs, and the number of distinct pairs (first JA, second JA) is variant. All of our synthetic tables have 2 columns, and the values are generated randomly. We also generated data from the TPC-DS benchmark [24] with a scaling factor 10 and then replicated the data 100 times.

In Table 1, the first and second JAs are defined according to the join queries. Note that table sizes run in the billions of rows, which is a good reason for replacing tables with learned models.

We analyze performance for the following four join queries, which are sufficient to reveal the key features:

Q1(SynthDB2): $tbl2 \bowtie tbl3 \bowtie tbl0$
Q2(SynthDB2): $tbl0 \bowtie tbl1 \bowtie tbl2 \bowtie tbl3 \bowtie tbl4$
Q3(TPC-DS): $Web\_sales \bowtie_{cus} Store\_sales \bowtie_{items} Store\_returns$
Q4(TPC-DS): $Ship\_mode \bowtie_{SM\_id} Web\_sales \bowtie_{SM\_id} Catalog\_sales$
$\bowtie_{items} Store\_sales \bowtie_{store\_id} Store\_returns \bowtie_{reason\_id} Reasons$

With synthetic data we test the worst case when there is no meaningful relations between the first and second JAs values. Thus, their distinct values are selected randomly.

**Reproducibility.** Table 5 shows the hyper parameters for models per table involved in the join queries. The third column shows the number of clusters, the fourth shows the number of nodes per layer (in all cases we use 5 layers), and the last shows the maximum epochs. AdamOptimizer is used for all models with learning rate 0.0005. Implementation used Python and Tensorflow. The code and documentation can be found at: https://github.com/shanghoosh1/ModelJoin

### 5.2 Efficiency and Overheads

We first illustrate the results of learning per-table models. Then, given the mdoels, we evaluate *Model Join*.

Learning the models consists of embeddings, clustering and training NNs. Table 2 shows the learning time costs for the joined tables in each query. Please note that these costs occur once and are reasonable, taking from a few minutes to several hours. Table 3 shows total storage costs for embeddings, clustering, un-conditional frequencies for JAs, and CDG NN. These run from a few MBs to a few hundred MBs even for tables of size of 100s of GBs.

Table 4 shows sample-generation times. The time cost in Table 4 varies from ca. 14 minutes to ca. 112 minutes to generate a sample with 100k rows. As each sample data point is generated independently from others, sample generation is embarrassingly parallelizable, so times will be decreased by $N$ with $N$ cores.

### 5.3 Quality of Models and Join Sample

*5.3.1 Per-table Model Accuracy.* Table 6 shows the accuracy of per-table models. Note that F-score is already very high. This is one of the reasons why tables can be "absent" safely. Note: in these experiments we purposely do not allow models to be 'too' accurate (keeping the number of the clusters low) because we wish to see exactly the effect of errors in the samples.

*5.3.2 KS-test and F-score of Generated Samples* The KS-test calculates the maximum difference between the CDF of distinct data points in the exact join result and the CDF of distinct data points in our approximate sample. It reveals whether the sample is a uniform sample of the exact join result (skeleton). The null hypothesis is that the sample is a uniform sample. KS is calculated as follows:

$$D_{n,m} = Sup_x |F_{1,n} - F_{2,n}| \quad (9)$$

where $F_{1,n}$ and $F_{2,m}$ are the CDF functions of the first and the second sample. $Sup_x$ is the supremum function. With significant level $\alpha$ the null hypothesis is rejected if $D_{n,m} > C(\alpha)\sqrt{\frac{n+m}{n \times m}}$, and $C(\alpha)$ is calculated from $\sqrt{-\frac{1}{2}ln\frac{\alpha}{2}}$ which gives us the critical value. If $D_{n,m}$ is less than the critical value, the null hypothesis holds (i.e., the sample is a uniform sample of the exact join). In this case $P(D_{n,m} > C(\alpha))$ is equal to the significance level. But, if not uniform, KS fails to show how far the sample is from uniformity. Enter the F-score

### Table 1: Data characteristics

| DB name | table name | NVDs in 1st JA | NVDs in 2nd JA | 1st_2nd Distincts | Num. of records |
|---|---|---|---|---|---|
| SynthDB1 | tbl0 | 100,000 | 100,000 | 100,000 | 149,895,600 |
| | tbl1 | 100,000 | 100000 | 500,000 | 749,967,400 |
| | tbl2 | 100,000 | 100000 | 1,000,000 | 1,499,818,300 |
| | tbl3 | 100,000 | 100000 | 2,500,000 | 3,747,965,800 |
| | tbl4 | 100,000 | 100000 | 5,000,000 | 7,499,960,400 |
| | tbl5 | 100,000 | 100000 | 7,500,000 | 11,244,111,000 |
| SynthDB2 | tbl0 | 10,000 | 5000 | 1,000,000 | 991,047,100 |
| | tbl1 | 50,000 | 5000 | 1,000,000 | 998,403,000 |
| | tbl2 | 100,000 | 5000 | 1,000,000 | 1,284,291,300 |
| | tbl3 | 5,000 | 10000 | 1,000,000 | 988,931,000 |
| | tbl4 | 5,000 | 50000 | 1,000,000 | 998,533,100 |
| | tbl5 | 5,000 | 100000 | 1,000,000 | 1,379,785,400 |
| | tbl6 | 10,000 | 10000 | 1,000,000 | 995,563,500 |
| | tbl7 | 50,000 | 50000 | 1,000,000 | 1,000,400,900 |
| TPC_DS | Store_sales | customers=273,443 | items=54,000 | 13,745,062 | 1,375,167,200 |
| | Store_returns | stores=27 | reasons=39 | 1053 | 136,278,000 |
| | Store_sales | items=54,000 | stores=183,284 | 1,452,252 | 1,375,193,700 |
| | Catalog_sales | ship_mode=20 | items=54,000 | 1,064,990 | 716,351,500 |

### Table 2: Time cost in seconds

| DB name | table name | 1st Embedding | 2nd Embedding | Clustering | Training | Total |
|---|---|---|---|---|---|---|
| SynthDB1 | tbl0 (att0 → att1) | 385 | 336 | 74 | 1009 | 1804 |
| | tbl1 (att0 → att1) | 801 | 763 | 186 | 1939 | 3689 |
| | tbl2 (att0 → att1) | 1389 | 1185 | 177 | 3655 | 6406 |
| | tbl3 (att0 → att1) | 1562 | 1211 | 257 | 8595 | 11625 |
| | tbl4 (att0 → att1) | 1046 | 1069 | 345 | 27344 | 29804 |
| | tbl5 (att0 → att1) | 1190 | 763 | 434 | 41462 | 43849 |
| SynthDB2 | tbl0 (att0 → att1) | 415 | 614 | 20 | 1295 | 2344 |
| | tbl1 (att0 → att1) | 621 | 606 | 22 | 1341 | 2590 |
| | tbl2 (att0 → att1) | 809 | 826 | 30 | 1691 | 3356 |
| | tbl3 (att0 → att1) | 500 | 512 | 21 | 1398 | 2431 |
| | tbl4 (att0 → att1) | 608 | 607 | 61 | 2186 | 3462 |
| | tbl5 (att0 → att1) | 812 | 853 | 193 | 4001 | 5859 |
| | tbl6 (att0 → att1) | 596 | 437 | 22 | 1402 | 2457 |
| | tbl7 (att0 → att1) | 616 | 622 | 57 | 2203 | 3498 |
| TPC_DS | Store_sales(customers → items) | 834 | 793 | 354 | 24537 | 26518 |
| | Store_sales(items → store_id) | 116 | 167 | 26 | 575 | 884 |
| | Catalog_sales(SM_id → items) | 97 | 68 | 28 | 5754 | 5947 |
| | Store_returns(store_id → reason_id) | 33 | 40 | 3 | 33 | 109 |

### Table 3: Storage cost in MBs

| DB name | table name | Embedding | Clustering | Freq | CDG NN | Total |
|---|---|---|---|---|---|---|
| SynthDB1 | tbl0 (att0 → att1) | 70 | 2 | 2 | 379 | 453 |
| | tbl1 (att0 → att1) | 110 | 7 | 3 | 448 | 568 |
| | tbl2 (att0 → att1) | 111 | 11 | 3 | 473 | 598 |
| | tbl3 (att0 → att1) | 114 | 23 | 3 | 473 | 613 |
| | tbl4 (att0 → att1) | 116 | 40 | 3 | 473 | 632 |
| | tbl5 (att0 → att1) | 117 | 53 | 3 | 473 | 646 |
| SynthDB2 | tbl0 (att0 → att1) | 9 | 4 | 0.2 | 120 | 133.2 |
| | tbl1 (att0 → att1) | 31 | 9 | 0.7 | 120 | 160.7 |
| | tbl2 (att0 → att1) | 59 | 12 | 1 | 120 | 192 |
| | tbl3 (att0 → att1) | 9 | 2 | 0.2 | 133 | 144.2 |
| | tbl4 (att0 → att1) | 31 | 2 | 0.7 | 234 | 267.7 |
| | tbl5 (att0 → att1) | 59 | 2 | 1 | 361 | 423 |
| | tbl6 (att0 → att1) | 11 | 4 | 0.2 | 135 | 150.2 |
| | tbl7 (att0 → att1) | 56 | 9 | 1 | 234 | 300 |
| TPC_DS | Store_sales(customers → items) | 186 | 78 | 2 | 430 | 696 |
| | Store_sales(items → store_id) | 30 | 3.6 | 0.7 | 0.5 | 34.8 |
| | Catalog_sales(SM_id → items) | 6.3 | 0.005 | 0.7 | 24 | 31 |
| | Store_returns(store_id → reason_id) | 0.008 | 0.002 | 0.85 | 0.5 | 1.36 |

**Table 4: Time-cost for generating a uniform 100k sample**

| Query | Inference | Sampling | Total |
|---|---|---|---|
| Q1 | 45 | 799 | 844 |
| Q2 | 157 | 1032 | 1189 |
| Q3 | 4060 | 2646 | 6706 |
| Q4 | 35 | 3492 | 3527 |

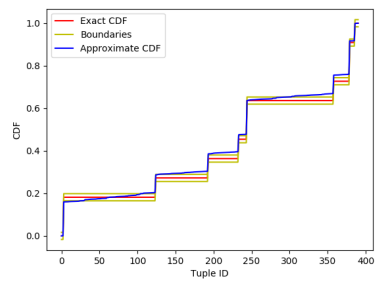

**Figure 7: CDF comparison for 7-way self join**

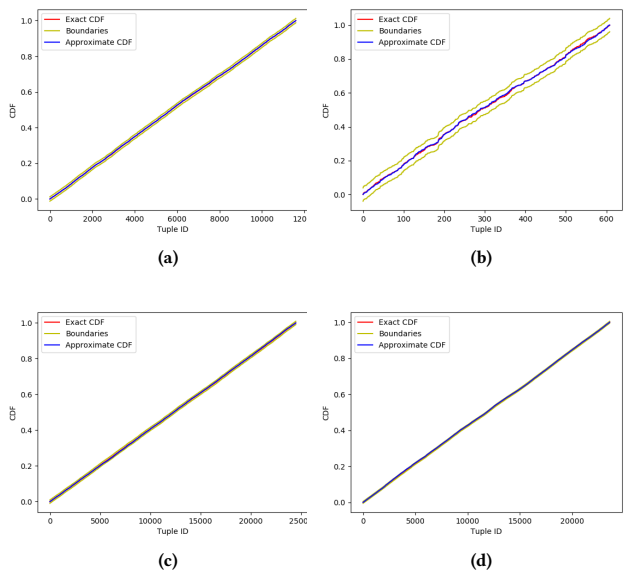

**Figure 6: a, b, c and d. CDF comparisons for Q1, Q2, Q3 and Q4 results**

metric, to shed light into the distance from uniformity, interpreting the quality of the approximate uniform sample.

Figure 6 shows the actual CDFs, the CDFs of our generated samples and the boundaries of KS-test that come from the critical values on TPC-DS queries. If our approximate CDF line goes outside the boundaries, it means the null hypothesis is rejected.

For Q1 and Q2, we made small tables from SynthDB2 with 20k samples, then built models from the first JA to the second JA in the tables. The exact join result sizes are 162,271 and 11,840 tuples respectively for Q1 and Q2. We generate samples with sizes 20k and 2k for Q1 and Q2, respectively. The critical values become 0.012 and 0.039. The KS-statistics for Q1, Q2 are 0.0029, 0.011. Thus the null hypothesis (easily) holds and the sample is declared uniform.

The F-scores for the sample generated for Q1 (Q2) are 0.87 (0.83).

For Q3 and Q4, again, we create smaller tables. For Q3 (Q4) we take 100k (5k) of the data points in the involved tables. Next, we build our models over them. The sizes of the exact join results of the small tables for Q3 and Q4 in the skeleton are 84,279 and 28,407,118. For Q3 and Q4, *Model Join* generates 50k and 100k, respectively. The KS critical values for these sample sizes are 0.0091 and 0.0051 respectively for Q3 and Q4 with $\alpha = 0.01$. The KS-statistic values are 0.0032 and 0.0048, respectively. These results imply that the null hypothesis (of the generated sample being uniform) is not rejected for both of them (see Figure 6).

Furthermore, the F-scores for Q3 (Q4) are 0.86 (0.81). Note, that (i) these scores can be made higher if more clusters are used; and (ii) such accuracy is actually good enough to ensure highly-accurate downstream LKD tasks, as we will show later.

### 5.4 Error Propagation in N-way Model Joins

How do individual model errors propagate in an n-way model join? To study this, we first compute (3-, 5- and 7-way) self joins of the same model with known error. We employ a synthetic table with 4k tuples and two attributes. The individual-model precision on the synthetic table is 0.97. The KS-statistics for the 3-, 5-, and 7-way self joins using significant level 1% are 0.006, 0.015 and 0.031, respectively. And the critical values are 0.02, 0.0177 and 0.0167, respectively. Thus, the 3- and 5-way join results are uniform. For the 7-way join, it is not uniform (the KS-statistic is larger than 0.0167 with significant level 1% - the CDF steps outside the boundaries in Figure 7).

To increase the interpretability of KS-test values, we use F-score as a complementary accuracy metric here. The size of the exact join results for 3-way, 5-way and 7-way joins are 15344, 54272 and 180224 respectively. We generate an approximate random sample of 10k tuples for each join. The F-score values for 3-way, 5-way and 7-way self joins are 0.945, 0.93 and 0.887 respectively.

Note that the error of uniformity in the 7-way self join can also to some extent be tolerable because all the downstream LKD tasks are approximate themselves. Otherwise, during the learning process, more clusters should be used to improve the accuracy.

### 5.5 Impact of Numbers of Clusters

Here, we use *SynthDB*1 tables. The relation between the pair of JA values here is random to stress-test the models and highlight the clustering benefits. Figure 8 shows F-scores on the 6 *SynthDB*2 tables (with 50 clusters). Each x-axis point refers to the number of distinct pairs (NDPs) in one of the 6 tables. The F-score worsens when increasing NDPs.

Figure 9 shows the positive effect on F-score when using more clusters with $M(tbl2)$ of synthDB1. Thus, when NDPs and NDVs in the first and second JAs increase and accuracy worsens, more clusters can improve F-score. Increasing the number of clusters, increases the number of models per table, but the size of each model per cluster is much smaller. Nonetheless, clustering helps us have accurate models. If more clusters are used, the smaller the cluster sizes will be, the higher the accuracy will be. Thus, we can use this knob to achieve high accuracy. In theory, the number of clusters could be equal to the number of distinct first-second JAs' pairs. Then, accuracy always will be perfect, but with a much higher cost.

Table 5: Hyper Parameters for training the models

| DB | Table | Clusters | Hidden_nodes | Max_epochs |
|---|---|---|---|---|
| SynthDB1 | tbl0 (att0 → att1) | 100 | 200 | 3 |
| | tbl1 (att0 → att1) | 100 | 200 | 3 |
| | tbl2 (att0 → att1) | 100 | 200 | 3 |
| | tbl3 (att0 → att1) | 100 | 200 | 3 |
| | tbl4 (att0 → att1) | 100 | 200 | 5 |
| | tbl5 (att0 → att1) | 100 | 200 | 5 |
| SynthDB2 | tbl0 (att0 → att1) | 50 | 200 | 3 |
| | tbl1 (att0 → att1) | 50 | 200 | 3 |
| | tbl2 (att0 → att1) | 50 | 200 | 3 |
| | tbl3 (att0 → att1) | 50 | 200 | 3 |
| | tbl4 (att0 → att1) | 50 | 200 | 3 |
| | tbl5 (att0 → att1) | 50 | 200 | 3 |
| | tbl6 (att0 → att1) | 50 | 200 | 3 |
| | tbl7 (att0 → att1) | 50 | 200 | 3 |
| TPC_DS | Store_sales(customers→ items) | 50 | 300 | 20 |
| | Store_sales(items→ store_id) | 5 | 10 | 10 |
| | Catalog_sales(SM_id→ items) | 20 | 30 | 10 |
| | Store_returns(store_id→ reason_id) | 5 | 10 | 10 |

Table 6: F-score and Confidence intervals with $\alpha = 95\%$

| DB name | table name | F-score | Intervals |
|---|---|---|---|
| SynthDB1 | tbl0 (att0 → att1) | 0.984 | 0.009 |
| | tbl1 (att0 → att1) | 0.9926 | 0.005 |
| | tbl2 (att0 → att1) | 0.9929 | 0.005 |
| | tbl3 (att0 → att1) | 0.9823 | 0.008 |
| | tbl4 (att0 → att1) | 0.948 | 0.013 |
| | tbl5 (att0 → att1) | 0.93 | 0.02 |
| SynthDB2 | tbl0 (att0 → att1) | 0.944 | 0.014 |
| | tbl1 (att0 → att1) | 0.968 | 0.01 |
| | tbl2 (att0 → att1) | 0.973 | 0.009 |
| | tbl3 (att0 → att1) | 0.922 | 0.16 |
| | tbl4 (att0 → att1) | 0.918 | 0.017 |
| | tbl5 (att0 → att1) | 0.93 | 0.015 |
| | tbl6 (att0 → att1) | 0.932 | 0.015 |
| | tbl7 (att0 → att1) | 0.955 | 0.13 |
| TPC_DS | Store_sales(customers→ tems)s | 0.941 | 0.014 |
| | Store_sales(items→ store_id) | 0.91 | 0.01 |
| | Catalog_sales(SM_id→ items) | 0.986 | 0.007 |
| | Store_returns(store_id → reason_id) | 0.99 | 0.005 |

### 5.6 Downstream LKD Over the Model Join Samples

We exemplify the usefulness of *Model Join* for downstream LKD tasks. Works in data analytics and management have already shown that uniform samples can be used to train accurate LKD models [8, 16, 17, 31]. Here we add to this by exemplifying that the uniform sample generated by *Model Join* can be used for classification tasks. We show results using the join result of Q4 for a binary classification task. The labels are the store-id (1 or 2) and the independent variables are all the other JAs in the result. The popular XGBClassifier from XGBoost library is used to learn the classification models with the same hyper parameter tuning.

First, we take the exact join result of Q4 (which is 7m rows) and split it into training (80%) and testing (20%) parts, randomly. Note, this testing set is used to test all the three scenarios below: (1) We train XGBClassifier with the training set (producing $\mathcal{M}_{main}$). (2)

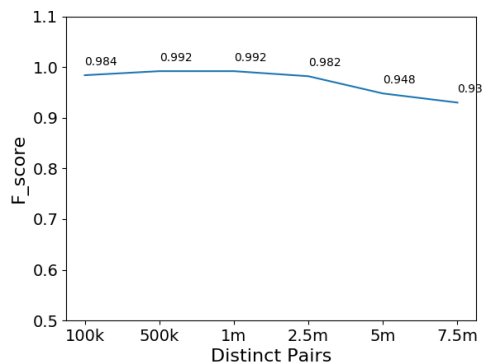

Figure 8: F-score vs. Number of Distinct Pairs

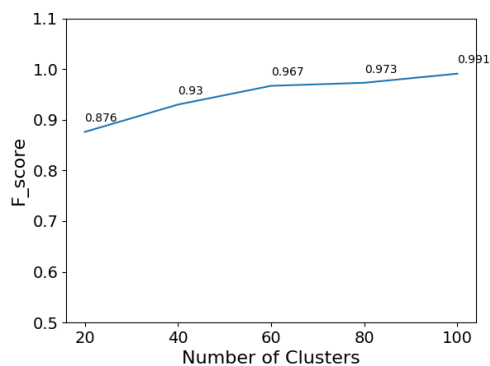

Figure 9: Effect of increasing the number of clusters

We create a 10k-sample from the training set (i.e., the 80% of the true join result) and train XGBClassifier on it (producing $\mathcal{M}_{sample}$). (3) we train XGBClassifier over the approximate 10k-sample produced by *Model Join* (producing $\mathcal{M}_{ModelJoin}$).

The F-scores for $\mathcal{M}_{main}$, $\mathcal{M}_{sample}$ and $\mathcal{M}_{ModelJoin}$ are 64.33, 62.89 and 62.71, respectively. This shows two facts: i) The F-scores for $\mathcal{M}_{sample}$ and $\mathcal{M}_{ModelJoin}$ are close to $\mathcal{M}_{main}$ which shows that using uniform samples from large tables instead of the raw tables for training is a good choice. In other words, the error due to sampling is small. ii) $\mathcal{M}_{sample}$ and $\mathcal{M}_{ModelJoin}$ have very similar F-scores. This shows that *Model Join* can indeed enable high-quality downstream LKD tasks, despite creating a sample of an approximated join result.

## 6 RELATED WORK

To the best of our knowledge there is no prior research for enabling LKD tasks over join results of absent tables. This work formulates and solves this problem.

**Join sampling when tables are not absent**. Joining the tables and then sampling defeats the point, as the join itself may be extremely time- and resource-consuming. For join sampling, without computing the join first, Olken [25] introduced a method based on rejection sampling to provide uniform and independent samples over joins. Chadhuri et al[4] also introduced a method to generate uniform and independent samples. These are applicable on two-way joins. [2] proposed join synopses, for foreign key joins. [26] suggests a factorized approach to learn regression models over the join result using factorization. This can only be used for specific tasks, such as regressions and aggregation function computation (i.e., it is not a *universal solution*). Zhao et al [34] extended Olken's and Chadhuri et al's approaches to produce uniform/independent samples for arbitrary n-way joins. In [22, 23] the authors introduced a more efficient way to generate the uniform samples of the join result. [27] generates weighted samples over stream joins. None of the above methods can be used over absent tables.

**Learning Models for Data Analytics**. ML models are increasingly being developed to improve accuracy/efficiency for analytical queries [8, 16, 17, 29]. Similarly, a large recent body of work has developed learned models for selectivity estimation [12, 30, 31]. These models are trained from uniform-independent samples and are examples of downstream LKD tasks performed on uniform samples, such as those enabled by this work.

The proposed learning framework generates (approximate) data tuples of the join result without any training data from the actual join result, nor from the joined tables. Instead, it uses learned models built and trained only on the individual to-be-joined tables. We show how to derive these per-table models. As PGMs are a natural fit for join graphs, we formulate our framework in terms of PGMs, the inference on which utilizes effectively the per-table models, which in turn consist of a novel blend of embeddings, clusterings, and generative neural networks.

## 7 CONCLUSION

LKD over the results of joining several relational tables is currently impossible if said tables are unavailable for any reason. This work addresses this important problem. We propose a new framework *Model Join*, which essentially can run the model join queries (which involves at least a model instead of tables). We formulate *Model Join* in terms of PGMs thereby creating a framework whereby per-table learned models can be 'plugged in' the PGM graph. With this, we can leverage PGM algorithms and a tweaked exact Variable Elimination Algorithm. We prove that if the models are 100% accurate, the framework does not introduce any additional error. The end result is a generative model, able to generate an arbitrary number of high-quality approximations of the data tuples in the join result. The output of the framework can be used as a *universal solution* in the sense that we do not single out any specific LKD task, but we wish to enable all. Experimental results with benchmark and synthetic data showed the high quality of the generated samples (using the KS-test for uniformity and F-score as metrics). It also quantified overheads and generation times, and exemplified high-quality downstream LKD tasks using the framework's output.

## 8 Acknowledgement

This work was supported by the UK Engineering and Physical Sciences Research Council (EPSRC) grant EP/R513374/1 for the University of Warwick.